\documentclass[aps,twocolumn]{revtex4}
\usepackage{graphicx,amsmath,amsfonts,amssymb, color}

\begin{document}

\title{Multifrequency and edge breathers in the discrete sine-Gordon system via subharmonic driving: theory, computation and experiment}  
\author{F. Palmero}
\affiliation{Grupo de F\'{\i}sica No Lineal.  Departamento de F\'{\i}sica Aplicada I.
Escuela T\'ecnica Superior de Ingenier\'{\i}a Inform\'atica, Universidad de Sevilla, Avda. Reina Mercedes, s/n, 41012-Sevilla, Spain}

\affiliation{Department of Physics and Astronomy  
Dickinson College, Carlisle, Pennsylvania, 17013, USA}
 
\author{J. Han}
\affiliation{Department of Physics and Astronomy  
Dickinson College, Carlisle, Pennsylvania, 17013, USA}

\author{L.Q. English}
\affiliation{Department of Physics and Astronomy  
Dickinson College, Carlisle, Pennsylvania, 17013, USA}

\author{T.J. Alexander}, 
\affiliation{School of Physical, Environmental and Mathematical Sciences, UNSW Canberra, Australia 2610}

\author{P.G.\ Kevrekidis}
\affiliation{Department of Mathematics and Statistics, University of Massachusetts,
Amherst MA 01003-4515, USA}

\affiliation{Center for Nonlinear Studies and Theoretical Division, Los Alamos
National Laboratory, Los Alamos, NM 87544}

\date{\today}

\begin{abstract}
We consider a chain of torsionally-coupled, planar pendula shaken horizontally by an external sinusoidal driver. It has been known that in such a 
system, theoretically modeled by the
discrete sine-Gordon equation, intrinsic localized modes, also known as discrete breathers, can exist. Recently, the existence of multifrequency 
breathers via subharmonic driving has been theoretically proposed and 
numerically illustrated by Xu {\em et al.} in Phys. Rev. E {\bf 90}, 042921 (2014). In this paper, we verify this prediction experimentally. Comparison of the 
experimental results to numerical simulations with realistic system parameters (including a Floquet stability analysis), and wherever possible
to analytical results (e.g. for the subharmonic response of the single driven-damped
pendulum), yields good agreement. Finally, we report on period-1 and multifrequency edge breathers which are localized at the open boundaries of the chain, 
for which we have again found good agreement between experiments and numerical computations.     
\end{abstract}

\maketitle

\section{Introduction}

Discrete breathers, also known as intrinsic localized modes, appear widely in damped-driven oscillator systems \cite{MarinN1996,AubryPD1997}, and general conditions for their appearance have been recently established theoretically \cite{HennigAA2013}.  Such time-periodic and exponentially localized
in space coherent structures have been observed experimentally in a diverse range of nonlinear oscillator systems, including Josephson junction arrays \cite{TriasPRL2000,BinderPRL2000}, coupled antiferromagnetic layers \cite{SchwarzPRL1999}, halide-bridged transition metal complexes~\cite{swanson},
micro-mechanical cantilever arrays \cite{SatoPRL2003,SatoC2003}, electrical transmission lines \cite{EnglishPRE2008} and  torsionally-coupled pendula \cite{cuevas} among others \cite{EisenbergPRL1998,EiermannPRL2004,BoechlerPRL2010}.  
They have also been argued to be of relevance to various biological
problems including dynamical models of the DNA double strand \cite{Peybi},
as well as more recently in protein loop propagation~\cite{niemi}.
Many of the features of the discrete breather response are generic across these wide-ranging experimental systems; see e.g. \cite{FlachPR2008}.
However, the intrinsic properties of a single oscillator (as well as,
often times, the specific nature of the coupling) may play a key role
in the observed dynamics and the nature of the discrete breathers
formed in the different physical systems.

Inspired by this observation, recent work has revealed that subharmonic resonances of a single oscillator (see e.g. \cite{Nayfeh1979}) may be used to excite discrete 
breather formation in an electrical lattice~\cite{EnglishPRL2012}.  
More recently, this idea has been examined further in the context of a horizontally shaken pendulum (which has long been known to display a variety 
of subharmonic resonances~\cite{StrubleQJMAM1965}), and the possibility of mixed-frequency breathers was identified in a pendulum chain~\cite{xu}.  These breathers exhibit 
the remarkable response that while energy is localized on a few pendula responding at a sub-harmonic of the driving force, the pendula in the tails of the breather 
are oscillating with the driving frequency.  
To the best of our knowledge, these theoretically proposed and 
numerically identified subharmonic breathers in the pendulum chain have
not yet been experimentally observed. This is one of the key goals
of the present work.  More specifically, we further investigate these mixed frequency breathers theoretically, and compute them numerically, exploring
their spectral and dynamical stability, identifying suitable
frequency intervals where they may be expected to persist. We then go on
to verify their existence by means of direct experimental observations in a 
horizontally shaken chain of 
torsionally-coupled pendula \cite{BasuThakurJPD2008,cuevas,english}. 

We also examine the role of breather location in the dynamics and reveal that discrete breathers may be localized at the end of the pendulum chain.  To the 
best of our knowledge this is the first time the existence of such
mechanical oscillator breather edge states has been experimentally 
demonstrated.
Nevertheless, it should be noted that 
research interest in edge states has a long history in other fields (see e.g. \cite{Davidson1996} and references therein), including manifestations
in the form of electronic surface waves at the edge of periodic 
crystals (Tamm states \cite{Davidson1996}), optical surface modes in 
waveguide arrays \cite{GaranovichPRL2008}, and more recently surface breather 
solitons in graphene nanoribbons \cite{SavinPRB2010}.  

Our presentation of the relevant results below is structured as follows.
In Section \ref{sec:model}, we present our theoretical model 
and discuss its physical parameters (of relevance to the experiment) 
for a horizontally shaken pendulum chain.  The relevant dynamical 
equation in the form of a damped-driven discrete sine-Gordon system is 
closely related to the driven-damped form of the famous 
Frenkel-Kontorova model~\cite{BraunPR2008,sgbook}.  
In Section \ref{sec:results}, after theoretically, numerically and
experimentally corroborating the subharmonic response of a single
pendulum, we seek subharmonic solutions numerically and trace their
parametric interval of stability. We are then able to show their
existence experimentally, both in the case of ``bulk'' subharmonic
breathers, as well as in the form of edge modes. Finally, in 
Section~\ref{sec:conclusions}, we summarize our findings and 
present some possible directions for future study.

\section{The model and experimental setup}
\label{sec:model}

The experimental setup is very similar to the one described in detail in Ref. \cite{english}. Each pendulum experiences four distinct 
torques - gravitational, torsional, frictional and driving torque. The driving torque arises due to the horizontal shaking of the pendulum 
array by a high-torque electric motor. The amplitude, $A$, of the sinusoidal driving was fixed in the experiment, but the frequency, $f = \omega_d/(2\pi)$, was 
finely tunable (in 0.05 Hz increments) and measured by magnetic sensing. Angles were measured using a horizontal laser beam from a diode laser attached 
to the frame of the pendulum array; this beam is then periodically interrupted by the swinging pendulum when properly aligned. This method gives an estimated 
precision of about $\pm 1 \deg$. An overhead web-cam was also used to monitor and record the pendulum motion.

As a result of the above contributions, the motion of a single 
(uncoupled) pendulum is well described by the equation,
\begin{equation}
\ddot{\theta} + \left(\frac{\gamma_1}{I}\right) \dot{\theta} + \omega_0^2 \sin\theta + F \omega_d^2 \cos(\omega_d t) \cos \theta = 0,
\label{eq:single}
\end{equation}  
where $I$ is the pendulum's moment of inertia, $I=ML^2+\frac{1}{3} mL^2$, the driving strength is given by $F=A \omega_0^2/g$, and $\omega_0$ is the pendulum's 
natural frequency of oscillation with $\omega_0^2=\frac{1}{I}(mgL/2+MgL)$. Experimentally, the number of pendula is $N=19$, $L=25.4$cm, 
$m=13$g, $M=14$g, $\gamma_1=500$ g cm$^2$/s, and $A=0.6$cm. Pendula at the two ends can oscillate freely (free boundary conditions).

If we add the torsional coupling to nearest-neighbor pendula, 
i.e., in the presence of all four of the above contributions,
Eq.~(\ref{eq:single}) becomes a system of differential equations given by,
\begin{eqnarray}
\ddot{\theta}_n + \omega_0^2 \sin\theta_n - \left(\frac{\beta}{I}\right) \Delta_2\theta_n + \left(\frac{\gamma_1}{I}\right) \dot{\theta}_n  & & \nonumber \\ 
-\frac{\gamma_2}{I}  \Delta_2\dot{\theta}_n +  F \omega_d^2 \cos(\omega_d t) \cos \theta_n & = & 0,
\label{eq:system}
\end{eqnarray} 
where $\beta$ is the torsional spring constant, and $\Delta_2$ represents the discrete Laplacian. We include an intersite friction term (prefactor $\gamma_2$) 
originating from 
the energy dissipation due to the twisting of the springs \cite{cuevas}.
Here, we assume that nonlinearity in the undriven array enters only through the sine-function in the gravitational term, but not through 
the coupling springs. This assumption seems to be experimentally justified for angle differences of up to 90 $\deg$, but it may not work well beyond that. Experimental values of 
coefficients are $\beta=0.0083$ Nm/rad and $\gamma_2=70$ g cm$^2$/s.
These equations can be non-dimensionalized by introducing the following parameters $\omega=\omega_d/\omega_0$, $C=\beta/I\omega_0^2$, $\alpha_1=\gamma_1/I\omega_0$,  
$\alpha_2=\gamma_2/I\omega_0$ and rescaling time $t \rightarrow t/\omega_0$, leading to the following dimensionless equation for the $n$th pendulum:
\begin{align}
\ddot{\theta}_n + \sin\theta_n - C \Delta_2\theta_n + \alpha_1 \dot{\theta}_n  & \nonumber \\ 
-\alpha_2  \Delta_2\dot{\theta}_n +  F \omega^2 \cos(\omega t) \cos \theta_n & =  0.
\label{eq:normsystem}
\end{align}
For our experimental conditions the dimensionless parameters 
are $C = 0.16$, $\alpha_1 = 64 \times 10^{-4}$, $\alpha_2 = 9 \times 10^{-4}$ and $F = 0.026$.    We use these 
parameters throughout the theoretical investigations of this work, and consider only variations in the dimensionless 
frequency parameter $\omega$, which is tunable as indicated above.  In our plots we transform back to physical units, plotting results versus 
driving frequency in Hertz, $f$, where, for reference, the natural frequency of our pendulum is $f_0 = \omega_0/(2\pi) = 1.04$Hz.
  
As numerical simulations have shown that a one-peak breather is mainly localized on a single pendulum and its 
first neighbors, experimentally, the method used to initiate multifrequency breathers is by manually displacing a group of three pendula through 
angles roughly predicted by the simulations. Upon release, a true breather mode can then sometimes establish itself, depending on whether the phase of release happened to be sufficiently close in relation to the driver. In practice, it may take a number of such trials before the driver can lock onto the initialized pendula in this manner. 

\section{Results}
\label{sec:results}

\begin{figure}
\includegraphics[width=3.0in]{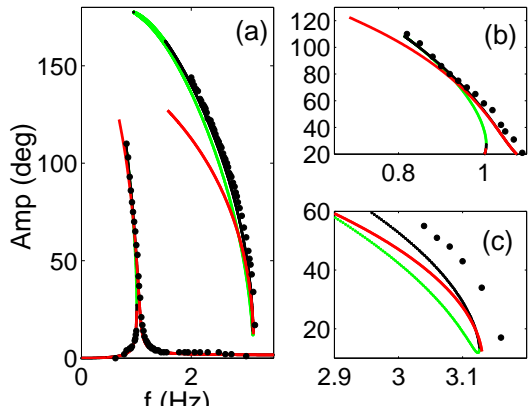}
\caption{(a) (Color online) The response curve of a single driven-damped pendulum. The filled circles indicate experimental data, the black and green lines are numerical results and the red line 
represents the analytical prediction. Black lines correspond to stable solutions and green to unstable solutions. For the main resonance at around 1 Hz, all three 
traces agree quite well. Notice, however, that due to the sine-expansion 
approximation, for the subharmonic resonance, the analytical prediction deviates from the numerical/experimental data for 
large amplitudes, as expected. (b) Zoom showing the peak corresponding to main resonance. (c) Zoom showing the origin of the subharmonic resonance.} 
\label{single}
\end{figure}

We first examine a single damped-driven pendulum. In general, we have observed  similar behavior to that found in \cite{xu}, where the same system was studied in a 
slightly different range of parameters.   Examining the response of the system to different frequencies and amplitudes of the driving force, we obtain the resonance curves shown in Fig. \ref{single}. Since a pendulum is an oscillator characterized by soft nonlinearity, we have found experimentally and numerically that the 
resonance curve exhibits the characteristic bend toward lower frequencies,
as is theoretically expected~\cite{Nayfeh1979}. 
At higher frequencies we find the well known pendulum subharmonic 
response~\cite{ChesterJIMA1975}.  A subharmonic branch starting at around three times the natural frequency can be obtained both in the 
experiment and in the numerics. Here, 
the pendulum responds to the driver by swinging at a frequency that is 
one-third of the 
driving frequency, $f$. In this way, for every three periods of the shaken table, the pendulum performs one complete swing. It is also 
interesting to note that larger response amplitudes can be achieved via subharmonic driving than with direct driving. Numerically we have found 
higher-order resonances, but these resonances correspond to frequencies not accessible in our experimental setup. In particular, we have 
found numerical solutions starting at around five and seven times the external driver frequency. Numerical simulations have shown that subharmonic
breathers corresponding to these high frequencies are mostly unstable, with the exception of frequencies within very narrow intervals close to the starting 
frequency value.

In order to get approximate analytical solutions to Eq.~(\ref{eq:single}), 
we Taylor-expand the trigonometric functions and obtain (in dimensionless form),
\begin{equation}
\ddot{\theta} + \theta +\left[\alpha_1 \dot{\theta} -\frac{\theta^3}{6}+ F \omega^2 \cos(\omega t)\left(1- \frac{\theta^2}{2}\right)\right] = 0,
\label{eq:single2}
\end{equation}

Assuming that in the main resonance case the solution takes the form, 
\begin{equation}
 \theta=V \cos(\omega t+ \phi),
\end{equation}
and in the subharmonic case,
\begin{eqnarray}
 \theta & = & V_{1/3} \cos(\omega t+\phi)+ V_{1/3} \cos(\omega t+ \phi_{1/3}) + \nonumber \\
 & & A_{1/3} \cos(\omega t/3)+ B_{1/3} \sin(\omega t/3),
\end{eqnarray}
and using the harmonic balance method \cite{Qaisia}, a set of algebraic equations can be deduced in order to get the values of parameters $V, \phi, V_{1/3}, \phi_{1/3}, A_{1/3}$ 
and $B_{1/3}$. Approximate resonance curves have been obtained, as shown in Fig. \ref{single}. We note that these approximate 
solutions show good agreement in the main resonance case 
(as previously also indicated in Ref.~\cite{xu}), 
but also in the subharmonic resonance case when
the amplitude oscillations are not too large. 
It is relevant to point out here that the Taylor expansion
utilized in order to obtain the analytical results is only valid
for small values of $\theta$; in that light, the range of agreement of
the theoretical results with the experimental (and numerical) ones
is well beyond the realm of applicability of the theoretical approximation.

\begin{figure}
\includegraphics[width=1.6in]{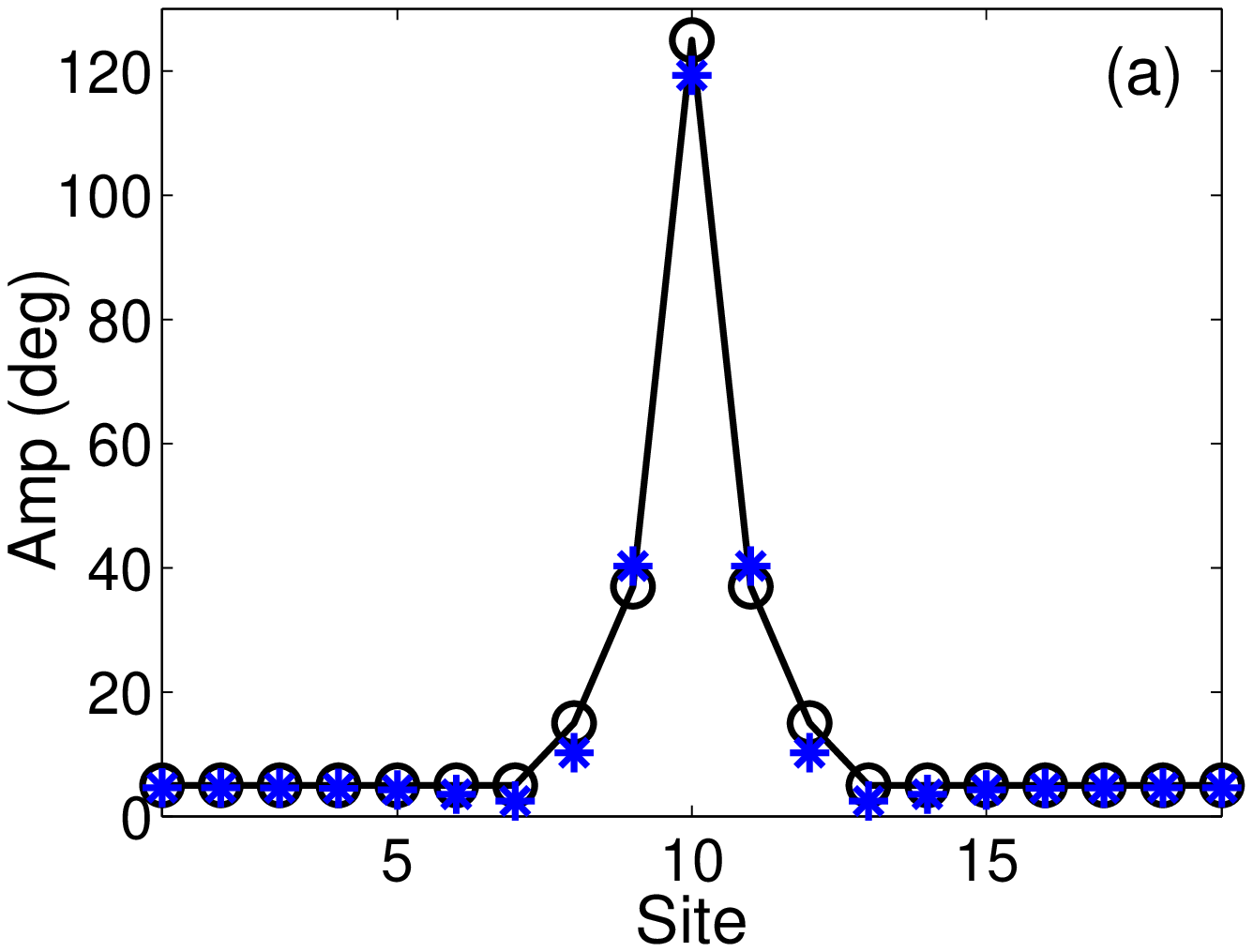}
\includegraphics[width=1.6in]{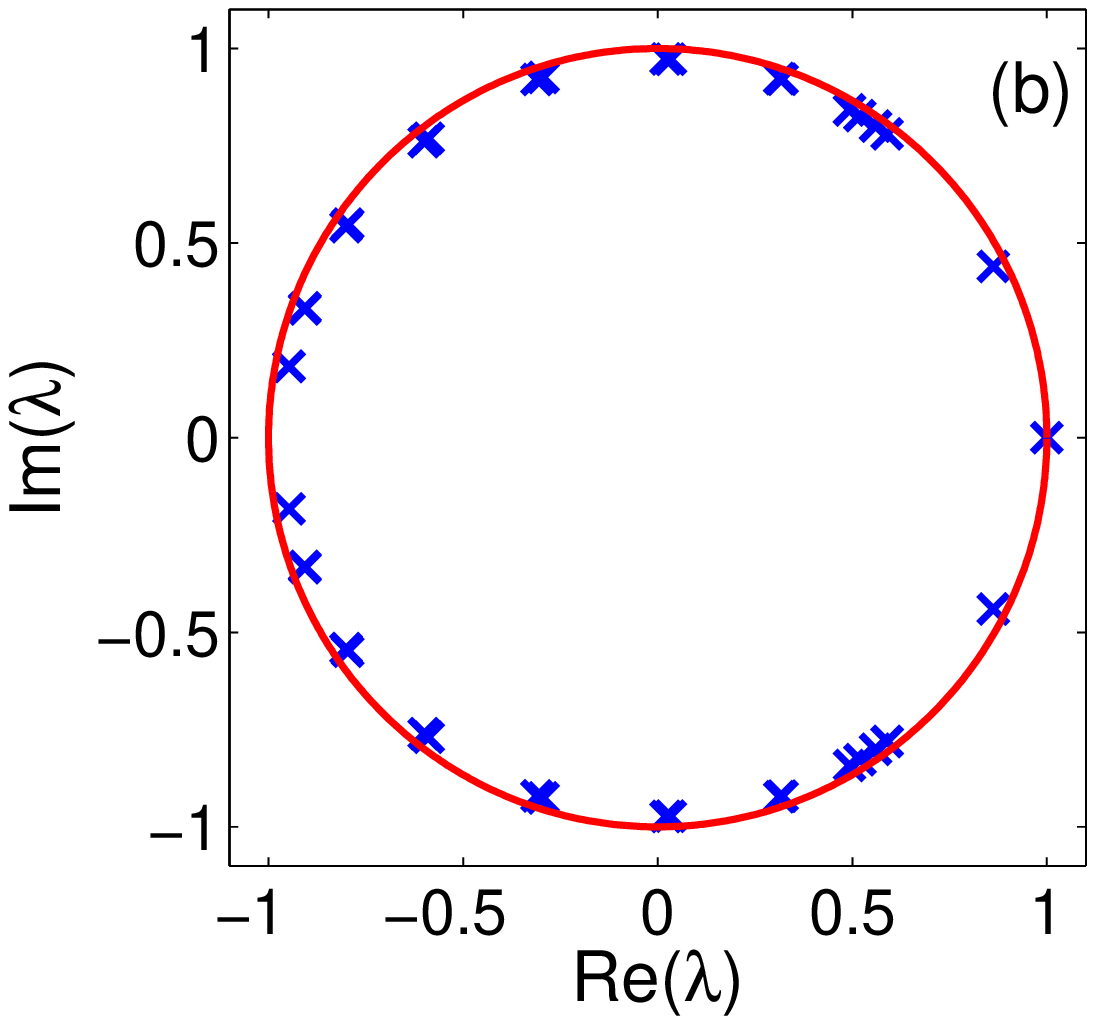}
\caption{The pendulum array: (a) The period-$1$ breather profile depicted as the maximum amplitude of each pendulum (no phase information). Experimental (numerical) angles are indicated by circles (stars). (b) The Floquet multipliers for the breather 
solution shown in the left panel are all within the unit circle, indicating 
its spectral stability. Both solutions correspond to a frequency of $0.91$ Hz. } 
\label{array0}
\end{figure}

Having mapped out the response regime for a single pendulum, let us now turn to the full pendulum array. The existence of period-1 breather solutions has already been established experimentally for this system \cite{english}. 
As a check, we start with the known period-1 breather and verify that experiments and numerics are in good 
agreement. This is demonstrated in Fig. \ref{array0} which depicts the maximum amplitude of oscillation for each pendulum. Numerical simulations 
performed with longer chains ($N=41$) show that this behavior is independent of the length of the chain.

\begin{figure}
\includegraphics[width=1.6in]{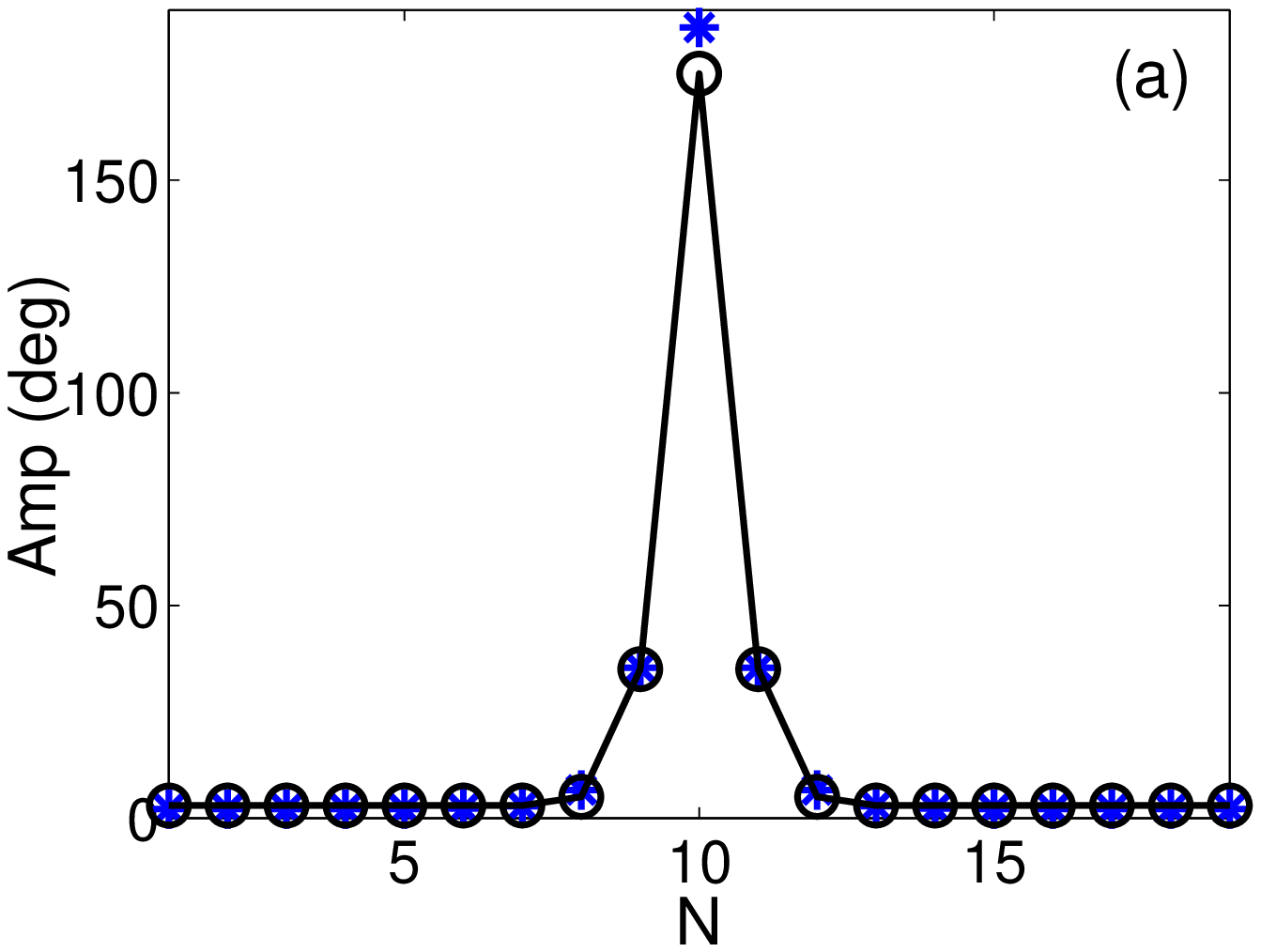}
\includegraphics[width=1.6in]{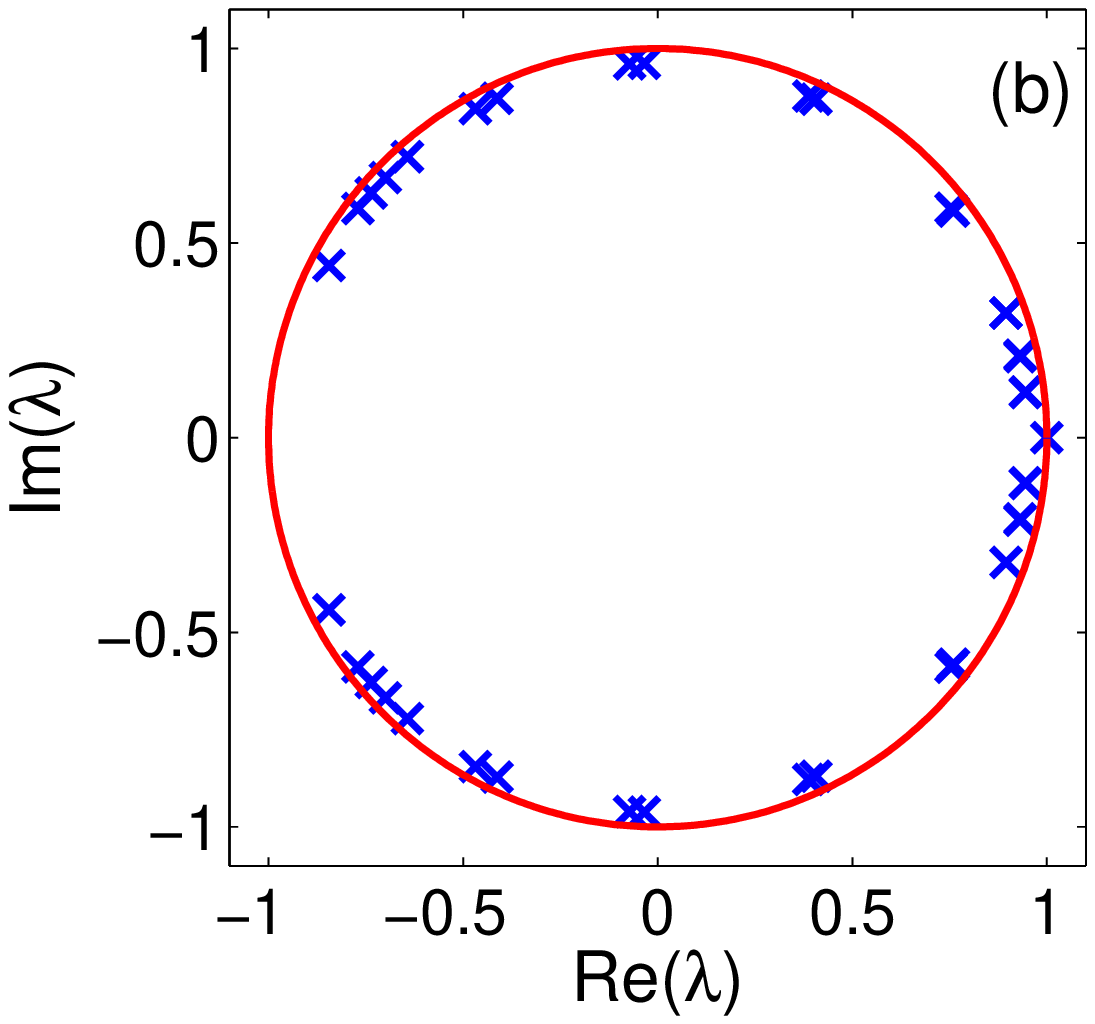}
\includegraphics[width=2.25in]{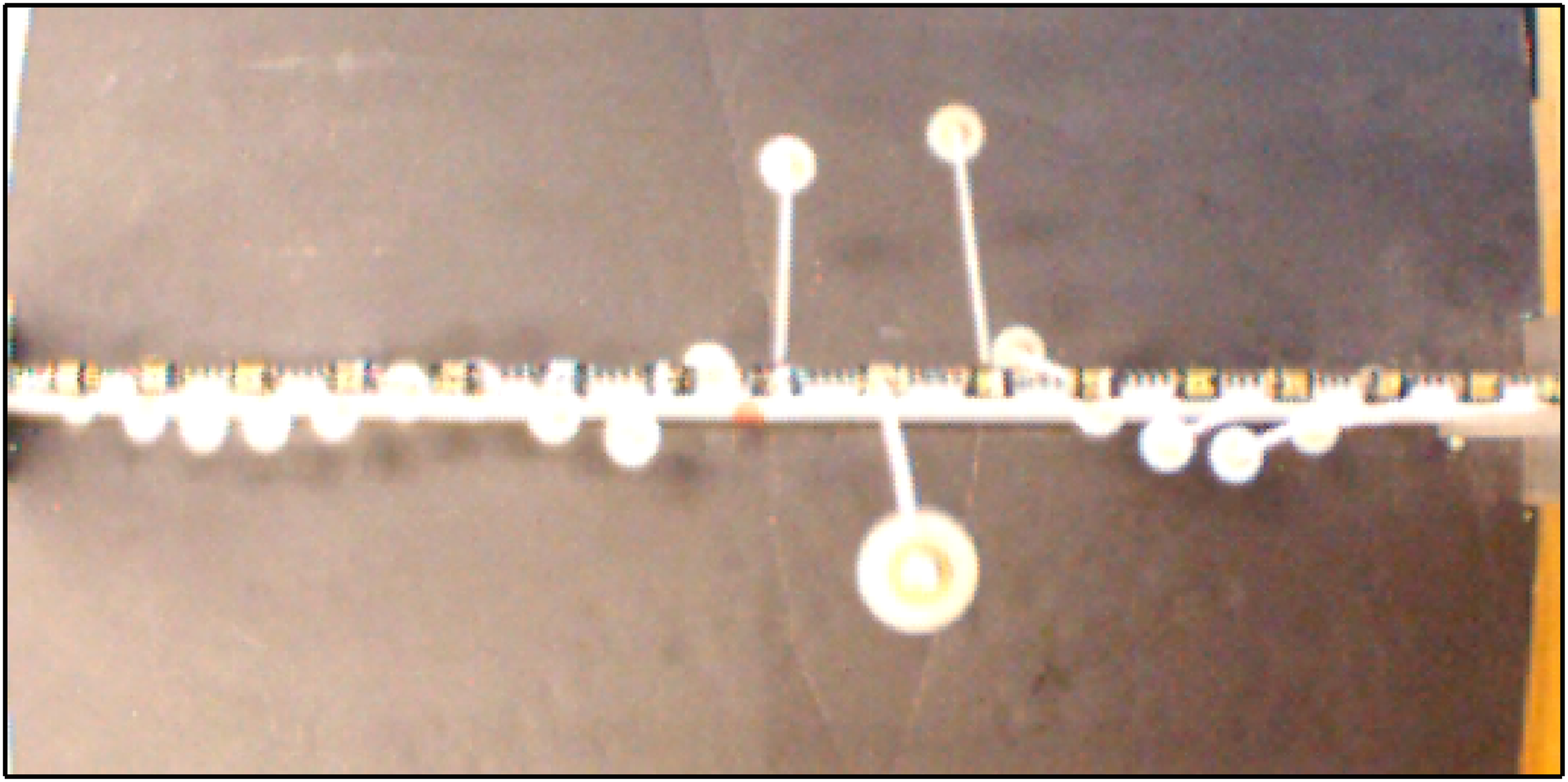}
\caption{The pendulum array:  (a) The subharmonic breather profile corresponding to 1.95 Hz depicted as the maximum amplitude of each pendulum (no phase information). Experimental (numerical) angles are indicated by circles (stars). (b) The Floquet multipliers for the 
breather solution shown in (a) are all within the unit circle, indicating breather stability. (c) The pendula as seen by the overhead 
web-cam. The multibreather mode at 1.75 Hz is sharply localized and the center pendulum is 
clearly seen to exceed 180 $\deg$.} 
\label{array1}
\end{figure}

Let us now consider the pendulum array in the case of 
the subharmonic response of the chain. One might expect that by turning on the coupling and moving away 
from the anti-continuous limit, and similar to the period-1 breather solution, 
a (multifrequency) breather emerges in which the center pendulum performs periodic (subharmonic) 
motion, whereas pendula in the wings of this mode respond weakly at the driving frequency. Previous numerical studies have shown that, for a different 
range of control parameter values, and close to the subharmonic bifurcation ($f\approx$ 3 Hz), this breather exists but is unstable, except for small frequency windows 
below the bifurcation point \cite{xu}. In our system, in contrast, we have been able to identify such a 
mode experimentally for a range of frequencies, and its existence and 
dynamical stability have been corroborated by numerical computations.
It should also be noted that, as indicated also in~\cite{xu}, the
precise stability details of such a 
subharmonic breather depend strongly on the number of pendula in the
chain, with a smaller $N$ favoring more robust configurations.

\begin{figure}
\includegraphics[width=3.0in]{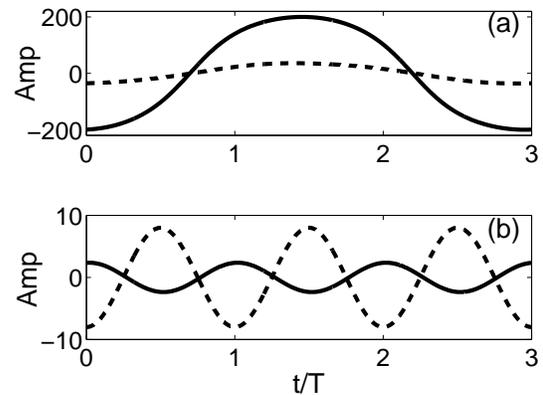}
\caption{Numerical mixed-frequency breather trajectory  across one period ($T=2 \pi/\omega$) for $f=1.95$ Hz and $N=19$ (amplitude in degrees). (a) Trajectory corresponding to  
central peak pendulum (continuous line) and its first neighbor (dashed line). (b) Trajectory 
corresponding to the edge pendulum at the end of the chain (continuous line) and normalized external driving force (dashed line), that is proportional to $-\cos(\omega_d t)$.}
\label{phase1}
\end{figure}

The mode profile corresponding to a frequency close to 2 Hz is mapped 
out in Fig. \ref{array1}(a). The x-axis denotes the
node index, and the y-axis plots the angles (away from vertically down) at the instantaneous turning point of the center pendulum. Note the excellent agreement 
between experiment and simulations in the left panel at around 1.95 Hz. In both traces, the center 
pendulum oscillates between roughly 180$^\circ$ and -180$^\circ$. In further agreement with numerical results, the peak of the experimental breather is observed to 
be out-of-phase with the tails at the turn-around points of the center pendulum (i.e., an out-of-phase breather), as shown in Fig. \ref{phase1}. (It should be noted that the angles of only the breather center and pendula to one side of it were experimentally measured to high precission, but the the breathers were visually found to be very close to symmetric about the ILM-center.)  

Fig. \ref{array1}(b) shows the Floquet multipliers of this solution demonstrating its stability.  Numerical simulations 
performed with longer chains ($N=41$) show that this behavior is independent on the length of the chain. Figure \ref{array1}(c) shows a snapshot recorded by the overhead camera for a driver frequency of 1.75 Hz. The mode is 
sharply localized with one pendulum acquiring an amplitude exceeding 180$^\circ$.

\begin{figure}
\includegraphics[width=3.0in]{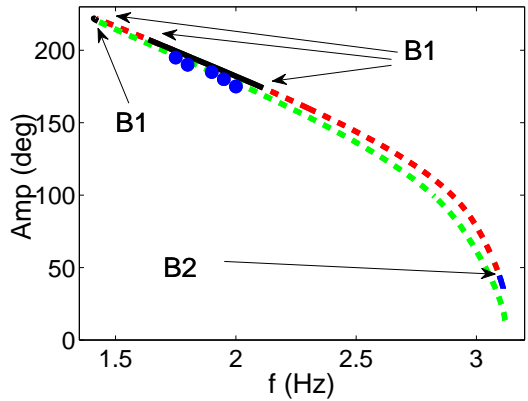}
\caption{(Color online) Amplitude in degrees plotted against the driver frequency corresponding to subharmonic breather families. The continuous black
and blue lines  indicate 
the two different regimes of stable numerical breathers (around $2$ and $3$Hz,
respectively). 
The dotted red lines show the
unstable breathers. The (blue) circles depict the experimental results. Notice the two frequency intervals of stability - one around 2 Hz, and the other, very small, around 3 Hz. The different bifurcations (B1: Bifurcation  associated with a conjugate pair of complex Floquet multipliers crossing the unit circle bifurcation; 
B2: Bifurcation  associated with a real Floquet multiplier crossing the unit circle) lead to the destabilization
of the breathers. The relevant scenarios are discussed in detail in the text.}
\label{array2}
\end{figure}

It is illuminating to study the effect of the driver frequency  on the profile of this multifrequency breather. Figure \ref{array2} maps out the 
amplitude of this breather solution as a function of the frequency; it thus represents a response curve for the multifrequency 
breather. We see (blue stars) that over 
much of the frequency interval that exhibited subharmonic response in the single pendulum, the breather solution is unstable against perturbations. There is, however, one 
band around 2 Hz and another narrow band around 3 Hz in which the multifrequency breather is predicted to be stable. Around 2 Hz, the stable breather is out-of-phase, by 
which we mean that the center and the tails are out of phase. At 3 Hz,  the stable breathers are more spread along the 
chain and the subharmonic frequency component (originating from the center pendulum) is still 
somewhat present at the tails. 

Our detailed numerical continuation for the experimental parameters
while varying the frequency identify two different stable subharmonic breather solutions - one at high frequency around 3 Hz and the other at low frequency around 2 Hz. Also note that there are two numerical solutions at each driver frequency (shown in Fig. \ref{array2} as red and green dotted lines), but the smaller-amplitude solution (green line) is always found to be unstable.
For the low-frequency family (around 2 Hz), the transition 
from stability to instability proceeds via a bifurcation
associated with a conjugate pair of complex Floquet multipliers crossing the
unit circle (B1), a Neimark-Sacker bifurcation (NSB). However, for the high-frequency breather family (around 3 Hz), the
relevant destabilization arises through a real multiplier
crossing the unit circle at $(1,0)$, as shown in detail in Fig. \ref{array2}. 

In our experiments, it has been possible to detect an interval of frequencies around 2 Hz where the subharmonic breather
exists, in good agreement with the numerical predictions. Also, in experiments, the NSB has been observed, occurring when a stable 
subharmonic breather, after a slight variation in the driver frequency, experiences oscillations that grow in amplitude until the breather finally vanishes.
This type of instability also occurs for the period-1 breather and was experimentally tracked and illustrated in Ref. \cite{BasuThakurJPD2008}.


\begin{figure}
\includegraphics[width=1.9in]{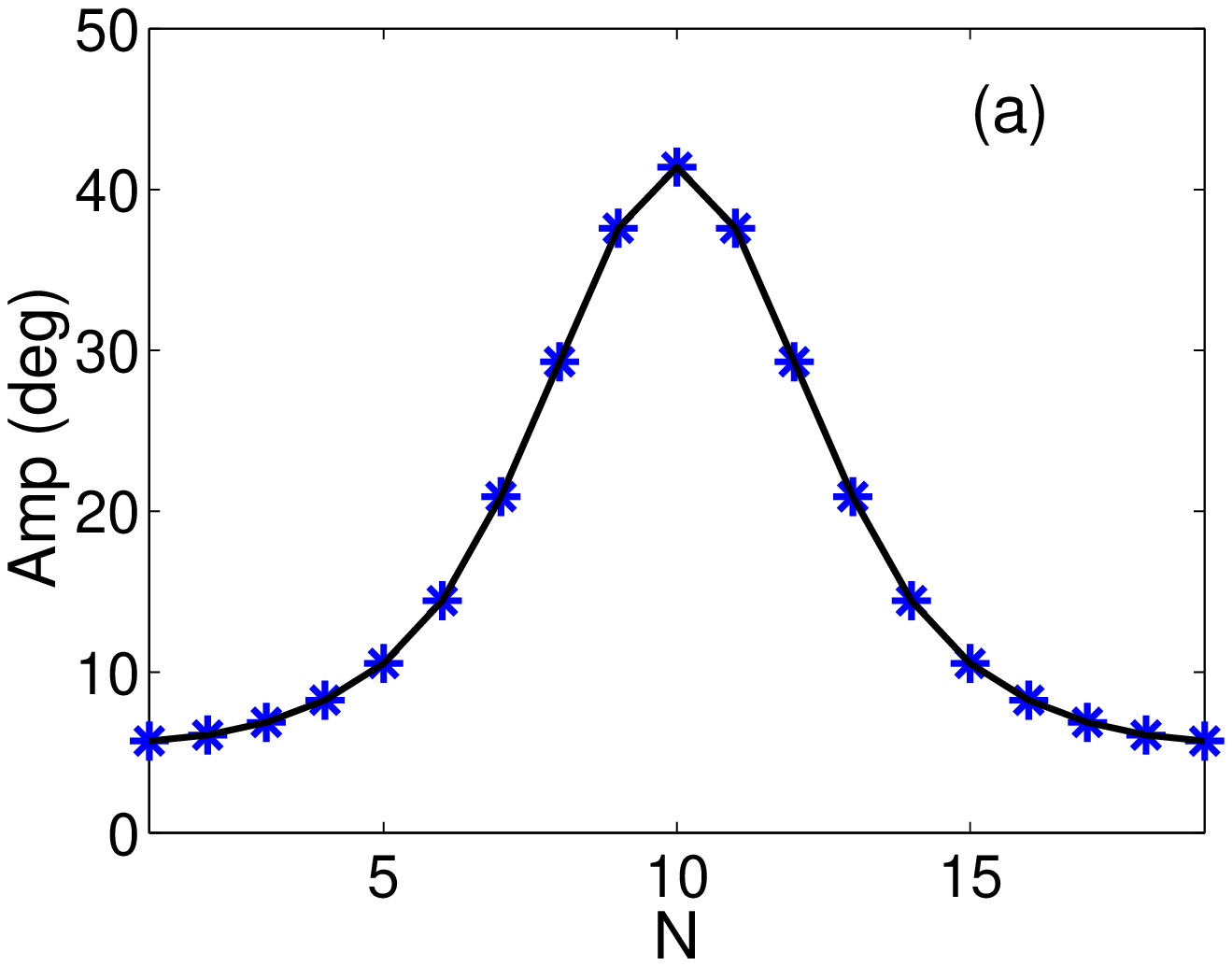}
\includegraphics[width=1.45in]{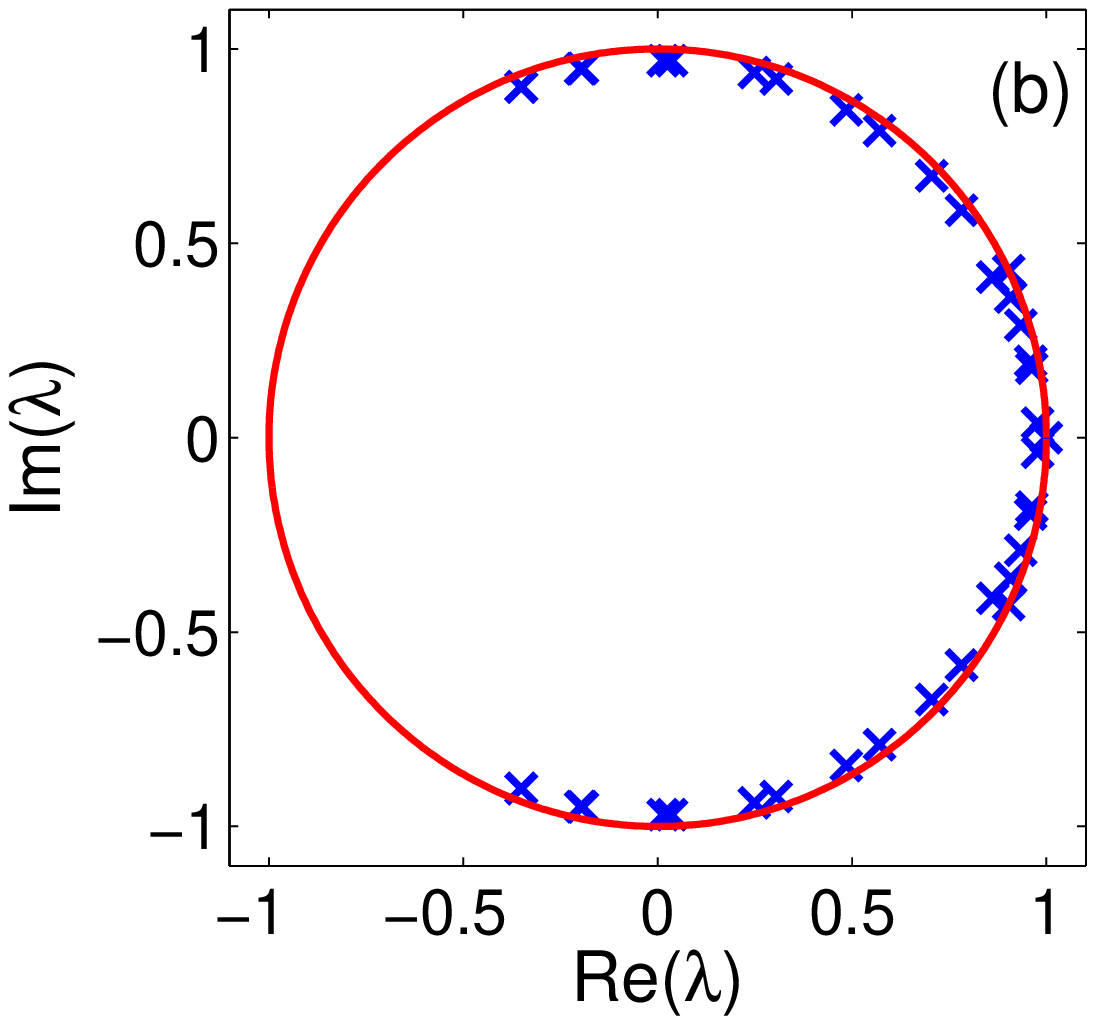}
\caption{Stable breather-like solutions corresponding to frequencies close to 3 Hz. (a) Breather profile (maximum amplitude in degrees) and (b) Floquet exponents corresponding to 3.1 Hz (one site breather).}
\label{array3}
\end{figure}

The question then arises if breathers can also be observed within the high-frequency interval (around 3 Hz). Numerically (and consonantly to the
above Floquet-multiplier stability analysis), we have 
found that around that value, in a very narrow interval of frequency values, different stable breather-like solutions can exist, very close in 
frequency values,  as shown in Fig. \ref{array3}. Simulations in longer chains show essentially the same phenomenon.  In experiments, in general, and for long 
time intervals, one-site and two-site breather-like transient states have been observed.

\begin{figure}
\includegraphics[width=1.9in]{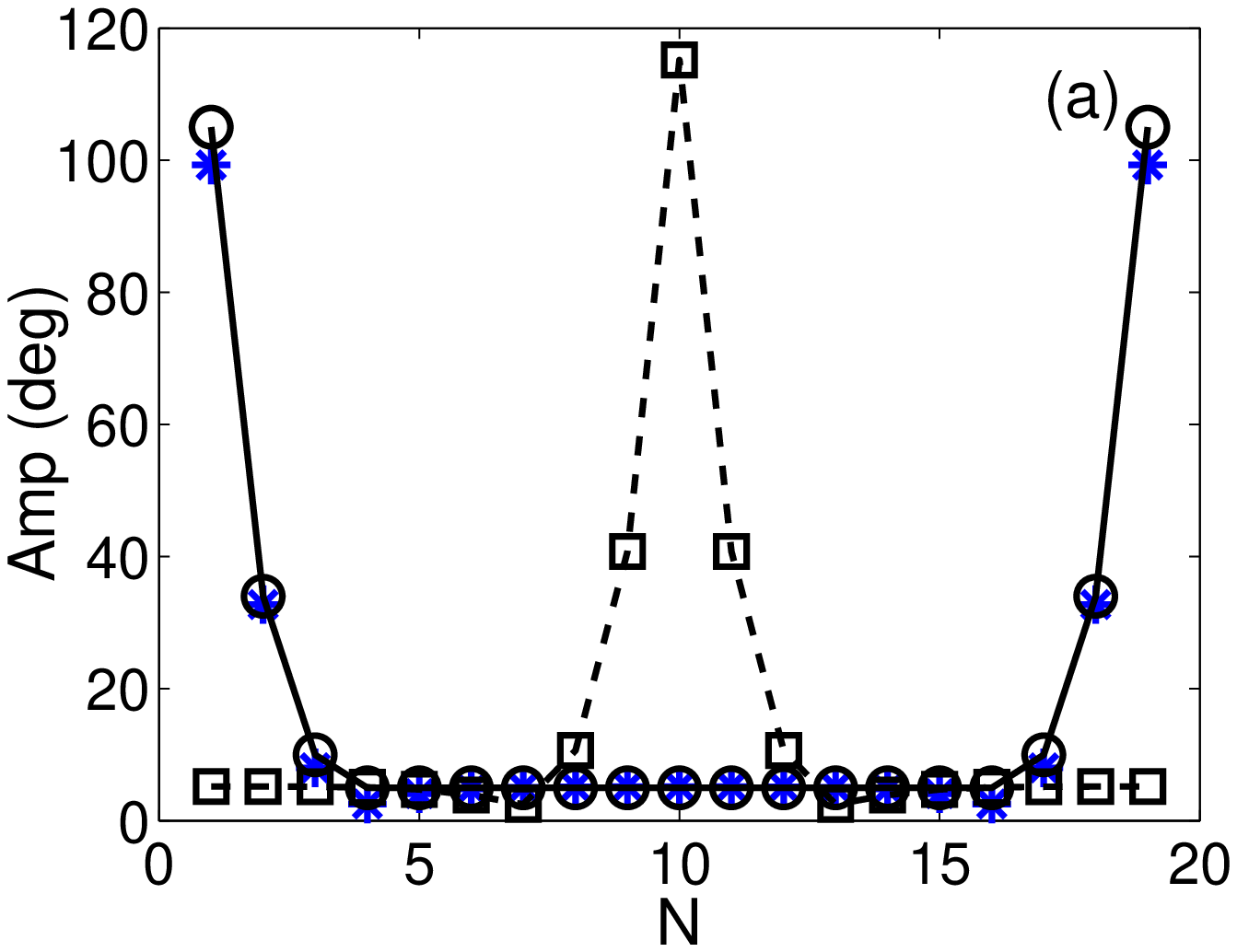}
\includegraphics[width=1.45in]{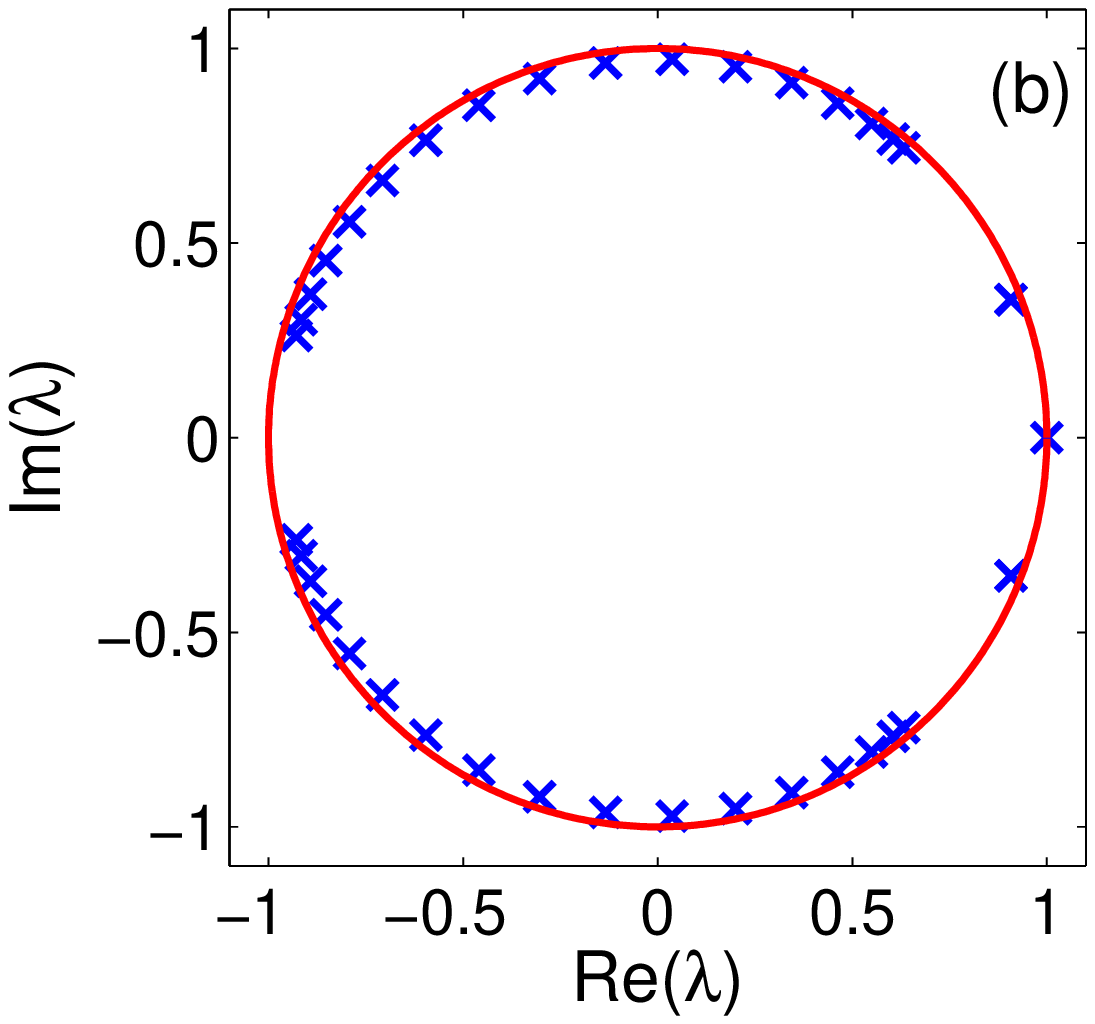}
\caption{A nonlinearly localized edge state for f = 0.92 Hz: (a) The edge breather profile - experimental (numerical) results are shown as circles (stars). Numerically 
stable breather located at the chain center and corresponding to the same frequency is shown as squares. (b) The Floquet multipliers for the edge breather are all within the unit circle indicating stability.} 
\label{edge_main}
\end{figure}

\begin{figure}
\includegraphics[width=3.0in]{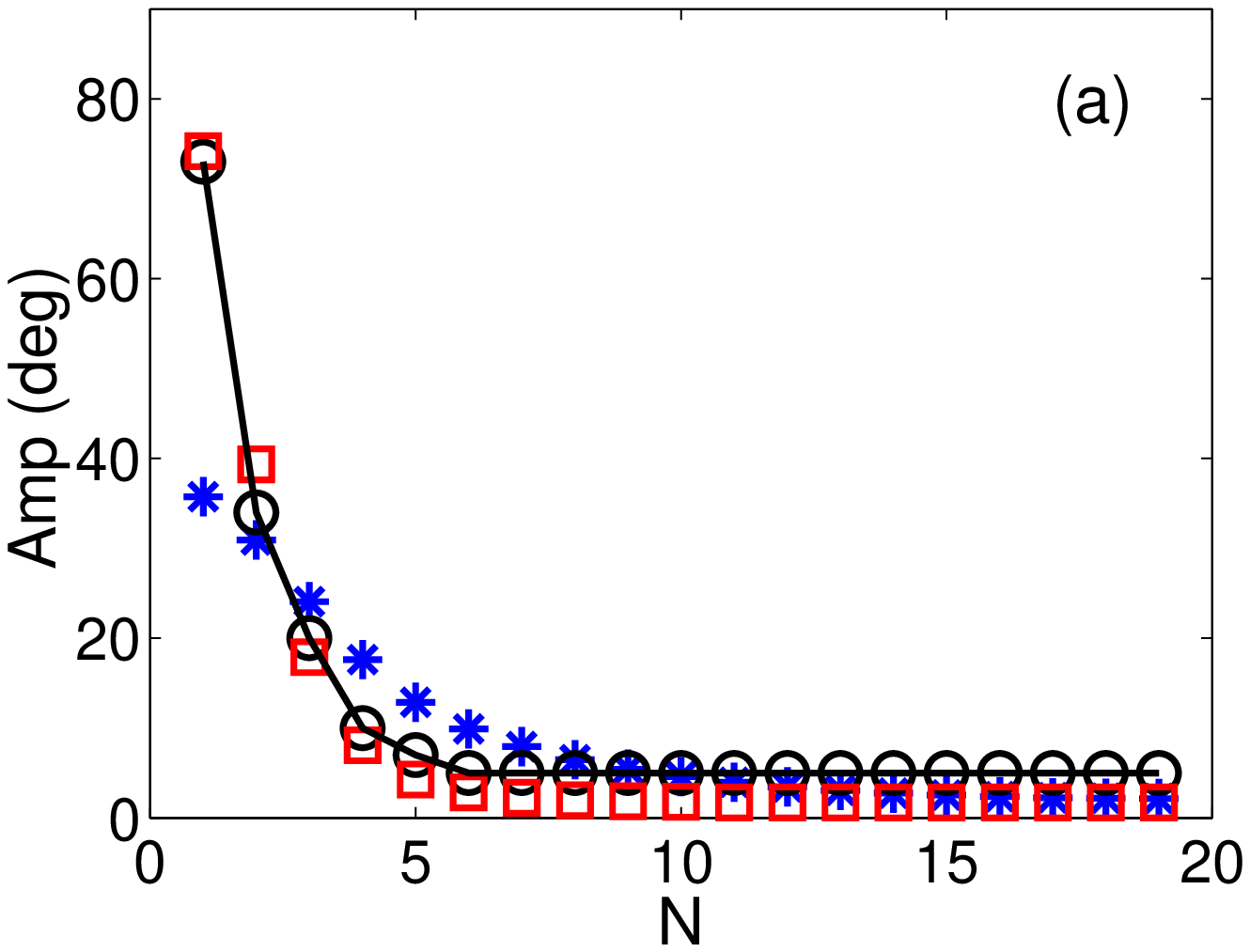}
\includegraphics[width=1.65in]{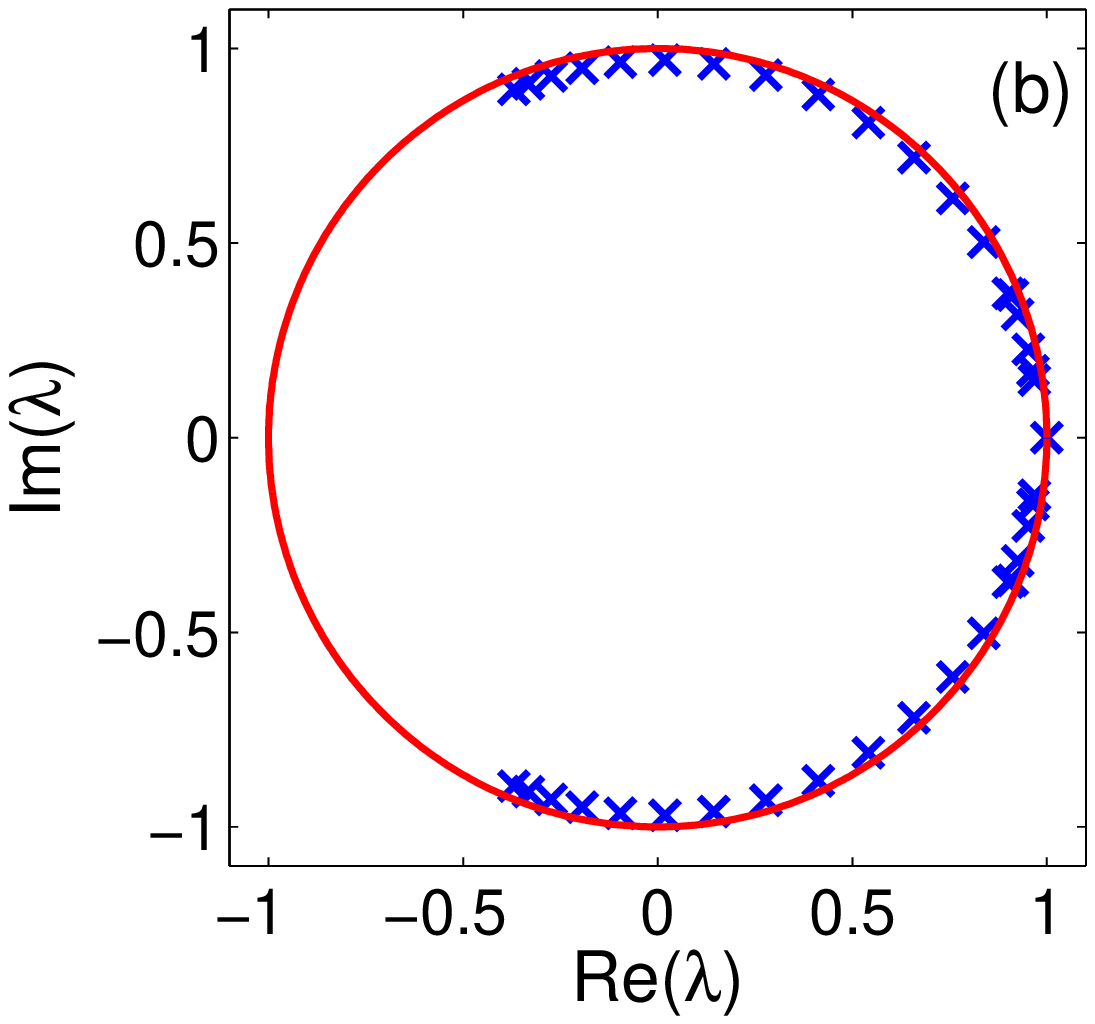}
\includegraphics[width=1.65in]{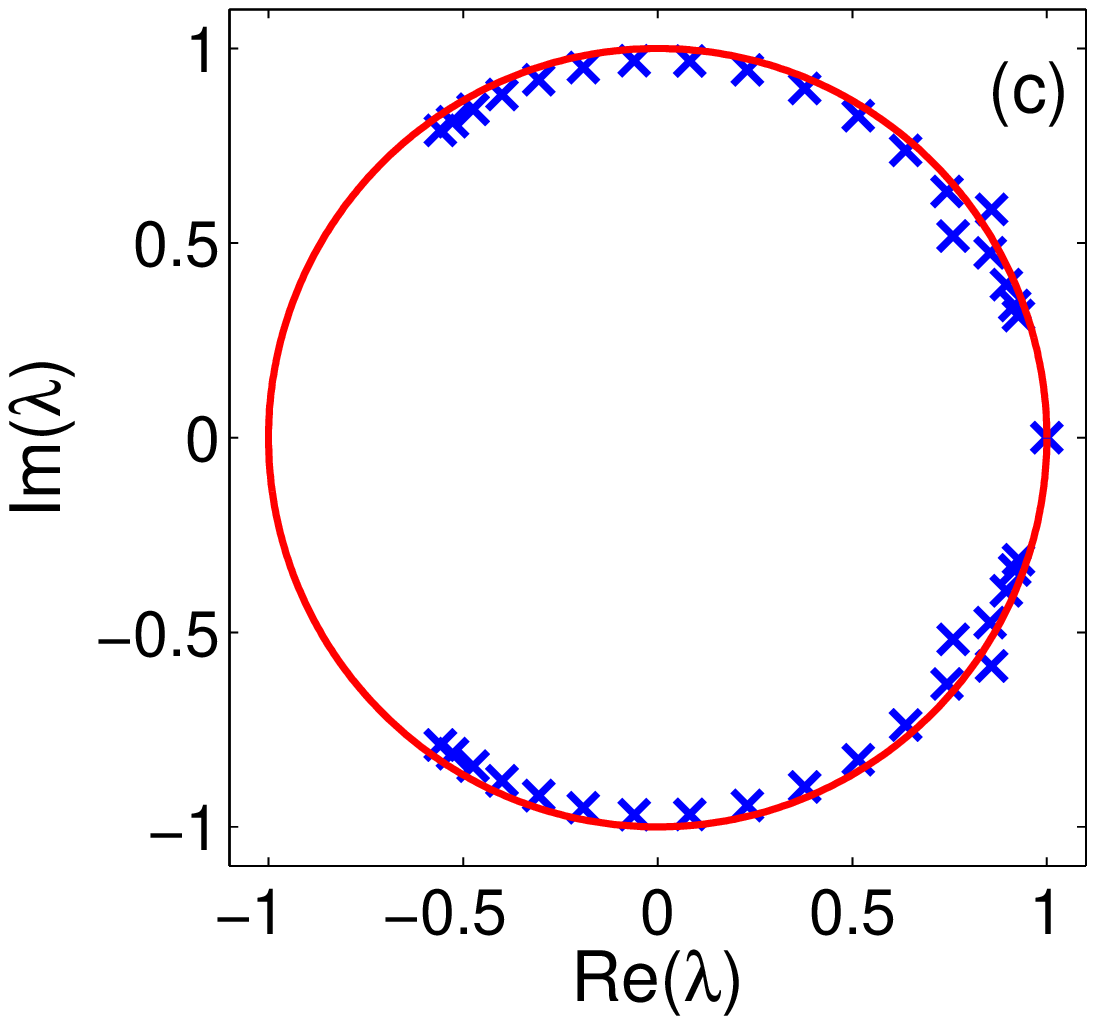}
\caption{A multifrequency edge breather driven subharmonically near f = 3.0 Hz. (a) The experimental profile  found for $f = 3.04$ Hz is depicted as black circles. In addition, two numerical solutions are shown: the red squares represent a 
solution  at $f = 2.97$ Hz that is found to be weakly unstable, see (c). The blue stars depict a nearby solution that is dynamically stable, see (b), corresponding 
to $f = 3.11$ Hz}. 
\label{edge_sub}
\end{figure}

We now turn to another type of nonlinearly localized mode that can be observed in the pendulum chain system, namely a mode that is 
localized at the chain boundary. 
Such modes have been extensively studied in other contexts such
as nonlinear optics~\cite{surface_m}, yet we are not aware
of such robust, experimentally demonstrated examples in pendulum
arrays. It should also be noted that such modes have also been recently
created in other damped-driven mechanical systems such as
e.g. granular crystals~\cite{hooge}.
As before, the two boundaries are open, and the driving is spatially homogeneous. Nonetheless, we can demonstrate both experimentally and numerically the existence (and stability) of modes 
localized over a few pendula near the edges of the chain with the interior pendula almost at rest. We find an edge state may be excited independently of the 
behavior at the other edge. An example with both edges excited is shown in Fig. \ref{edge_main}. The driver frequency (as well as the response frequency) here is chosen below the linear dispersion curve at $f=0.92$ Hz.  Note that the linear standing-wave modes are confined to the interval between 1.04 Hz and 1.34 Hz.  
The experimental data (circles) and the numerical simulation (stars) show close agreement - in both cases, the edge pendula attain an amplitude of 
oscillation of roughly 100$^\circ$, whereas the next pendula further in are found to be below 40$^\circ$ in amplitude. Floquet analysis demonstrates that the numerical profile is indeed stable, corroborating the experimental observability of
the relevant mode. 
We find that an edge breather typically has a lower maximum amplitude than the corresponding bulk breather, and it exhibits a more extended domain of stability.

As is evident in the numerical simulations, the amplitude of this edge-breather is frequency dependent, with lower driver frequencies giving rise to larger angles. It is clear that this mode should not be confused with a linear standing wave, where the edges also exhibit large oscillations. For instance, the second mode (just above the uniform mode) corresponding to a frequency of 1.048 Hz has the two pendula at the opposite edges oscillate $\pi$ out-of-phase. The nonlinear mode discussed here, however, oscillates in-phase and is sharply localized at the edges with interior pendula almost at 
rest. Furthermore, numerical simulations in longer chains show the same phenomenon.

It is relevant to point out that a subharmonic version of these edge-breathers also exists around a frequency, $f$, of 3 Hz. Figure \ref{edge_sub} presents the experimental data as 
black circles ($f=3.04$ Hz). We see that the edge pendulum swings with an amplitude of around 75$^\circ$, but crucially now at a third of the driver 
frequency, $f/3$. The response of the interior pendula, in contrast, is dominated by the driver frequency, $f$. This mode was found to persist throughout the time horizon of our experiments. 

Figure \ref{edge_sub} superimposes two numerical traces for two closely-spaced driver frequencies ($f=2.97$ Hz and $3.11$ Hz). The first one (squares) matches the 
experimental observation very well, but Floquet analysis reveals it to be weakly unstable. In fact, the instability is of the NS type, in contrast to the previous 
subharmonic breather (in the chain interior). The second numerical trace (stars) features a slightly larger $f$, and is stable, but departs from the experimental profile. It is 
likely that either small lattice inhomogeneities or weak nonlinearities in the torsional springs were responsible for stabilizing the observed multifrequency edge mode in the 
experiment.   

\section{Conclusions}
\label{sec:conclusions}

In the present work, we revisited the chain of coupled torsion pendula, an
experimental setup for which there is a well-established theoretical
model accounting for torsional contributions from gravity, nearest
neighbors, friction and external drive. This damped-driven system
was previously illustrated to feature prototypical discrete breather
waveforms. Recent explorations~\cite{xu} have suggested the possibility 
of subharmonic breather structures, which are unstable for
larger chains but potentially stable for smaller lattices.
In the present work, we have confirmed this expectation via
a combination of theoretical modeling (and where possible analysis),
numerical computation and experimental realization. We have indeed
observed not only period-1 breathers but also multifrequency / subharmonic
breathers as robust structures in the system, in line with our numerical
computations, in the appropriate parametric regimes guided by the
theoretical analysis. In addition to revealing the subharmonic breathers
and their good agreement with experimental observations, we have also
revealed surface breather modes, which have also been experimentally
identified.

Clearly, this system is a prototypical one for the exploration
of nonlinear structures, of their properties and interactions.
One can envision multiple directions for future investigation,
including the study of modulational instability (MI) as a source
for the generation of the breathers, the examination of 
multipeak breather structures (that have emerged as being relevant
here --due to their connection to the high-frequency branch--, but are
also a potential by-product of the MI), the study of breather-breather
or breather impurity interactions/scattering, among many others.
A number of these topics are presently under investigation and
will be reported in future publications.

\vspace{2mm}

{\it Acknowledgments.} F.P. acknowledges Dickinson College's hospitality and support. P.G.K.~gratefully acknowledges the support of
NSF-DMS-1312856, BSF-2010239, as well as from
the US-AFOSR under grant FA9550-12-1-0332,
and the ERC under FP7, Marie Curie Actions, People,
International Research Staff Exchange Scheme (IRSES-605096).
The work of PGK at Los Alamos is partially supported
by the US Department of Energy.

\end{document}